\documentclass[sigconf]{acmart}

\usepackage{multirow}
\usepackage{colortbl}
\usepackage{bm}
\usepackage{enumitem}
\usepackage{mathtools}

\usepackage{float}
\usepackage{listings}

\definecolor{codegreen}{RGB}{0,128,0}
\definecolor{codeblue}{RGB}{0,0,255}
\definecolor{backgray}{RGB}{248,248,248}
\definecolor{numgray}{gray}{0.4}

\lstdefinestyle{ExactPython}{
  language=Python,
  backgroundcolor=\color{backgray},
  basicstyle=\linespread{0.9}\ttfamily\footnotesize, % slightly tighter line spacing
  numbers=left,
  numberstyle=\tiny\color{numgray},
  numbersep=6pt,               % tighter number padding
  frame=single,
  rulecolor=\color{black},
  framerule=0.4pt,
  framesep=3pt,                % inner padding
  aboveskip=3pt,               % vertical spacing before
  belowskip=3pt,               % vertical spacing after
  columns=fullflexible,
  xleftmargin=1pt,             % subtle horizontal offset
  framexleftmargin=1pt,
  framexrightmargin=1pt,
  showstringspaces=false,
  tabsize=4,
  breaklines=true,
  keepspaces=true,
  commentstyle=\color{codegreen},
  keywordstyle=\color{codeblue},
  stringstyle=\color{black},
  emph={
    loaders.msmarco.load_queries,
    wrap_pyserini_searcher,
    from_prebuilt_index,
    load_queries,
    create_reformulator,
    reformulate_batch
  },
  emphstyle=\color{codeblue},
}

\renewcommand\footnotetextcopyrightpermission[1]{} % removes footnote with conference information in first column
\acmBooktitle{}

\begin{document}
%\texttt{Task Subspace Coverage}:
\title{QueryGym: A Toolkit for Reproducible LLM-Based Query Reformulation}

\author{Amin Bigdeli}
\authornote{Both authors contributed equally to this research.}
\orcid{0009-0003-8977-9312}
\affiliation{%
  \institution{University of Waterloo}
  \city{Waterloo}
  \state{Ontario}
  \country{Canada}
}
\email{abigdeli@uwaterloo.ca}

\author{Radin Hamidi Rad}
\authornotemark[1]
\orcid{0000-0002-9044-3723}

\affiliation{%
     \institution{Mila – Quebec AI Institute}
     \city{Montreal}
     \state{Quebec}
     \country{Canada}
}
\email{radin.hamidi-rad@mila.quebec}

\author{Mert Incesu}
\affiliation{%
  \institution{University of Toronto}
  \city{Toronto}
  \state{Ontario}
  \country{Canada}
}
\email{mert.incesu03@gmail.com}

\author{Negar Arabzadeh}
\orcid{0000-0002-4411-7089}
\affiliation{%
  \institution{University of California, Berkeley}
  \city{Berkeley}
  \state{California}
  \country{USA}
}
\email{negara@berkeley.edu}

\author{Charles L. A. Clarke}
\orcid{0000-0001-8178-9194}
\affiliation{%
  \institution{University of Waterloo}
  \city{Waterloo}
  \state{Ontario}
  \country{Canada}
}
\email{claclark@gmail.com}

\author{Ebrahim	Bagheri}
\orcid{0000-0002-5148-6237}
\affiliation{%
  \institution{University of Toronto}
  \city{Toronto}
  \state{Ontario}
  \country{Canada}
}
\email{ebrahim.bagheri@utoronto.ca}

\begin{abstract}
We present \texttt{QueryGym}, a lightweight, extensible Python toolkit that supports large language model (LLM)-based query reformulation. This is an important tool development since recent work on llm-based query reformulation has shown notable increase in retrieval effectiveness. However, while different authors have sporadically shared the implementation of their methods, there is no unified toolkit that provides a consistent implementation of such methods, which hinders fair comparison, rapid experimentation, consistent benchmarking and reliable deployment. \texttt{QueryGym} addresses this gap by providing a unified framework for implementing, executing, and comparing llm-based reformulation methods. The toolkit offers: (1) a Python API for applying diverse LLM-based methods, (2) a retrieval-agnostic interface supporting integration with backends such as Pyserini and PyTerrier, (3) a centralized prompt management system with versioning and metadata tracking, (4) built-in support for benchmarks like BEIR and MS MARCO, and (5) a completely open-source extensible implementation available to all researchers. \texttt{QueryGym} is publicly available at \url{https://github.com/radinhamidi/QueryGym}.
\end{abstract}

\maketitle
\section{Introduction}

Query reformulation and expansion play a central role in Information Retrieval (IR), particularly in scenarios where the initial user query is underspecified, ambiguous, or contextually sparse \cite{rocchio1971relevance,abdul2004umass,lavrenko2017relevance,penha2022evaluating}. Building on recent advances in large language models (LLMs), a growing body of work has introduced techniques that employ LLMs to generate enriched or contextualized variants of user queries, with the goal of improving alignment between query intent and relevant documents \cite{query2doc,Genqrensemble,Genqr,LameR}. These approaches frequently demonstrate strong retrieval gains in zero-shot and few-shot settings, largely due to their reliance on prompt-based generation rather than supervised training \cite{exploringresearch,qa-expand,mugi,csqe}.

Despite increasing interest in LLM-driven query expansion, progress in this area is constrained by the absence of a dedicated, reusable software framework that enables systematic development and reproducible experimentation. Existing methods~\cite{mugi,csqe,query2doc,LameR} present three recurring limitations. First, many approaches lack publicly released implementations~\cite{query2doc,exploringresearch,LameR,Genqr}, and the limited codebases that are available are often tightly bound to specific datasets, prompt templates, or retrieval backends, reducing their applicability across benchmarks and domains \cite{mugi,csqe}. Second, current implementations typically do not provide standardized interfaces, modular components, or extensible system abstractions. As a result, adapting these methods to new datasets, modifying prompting strategies, or integrating with alternative retrieval pipelines requires substantial engineering overhead. Third, reproducibility remains difficult due to undocumented dependencies, hardcoded configurations and prompts, ad hoc scripts, and inconsistent output formats. These challenges, among others, necessitate a unified and extensible toolkit that would facilitate rapid experimentation, systematic assessment of prompt and model variations, and ensures reproducible work in LLM-based query reformulation.

\begin{figure*}[t]
\centering
\includegraphics[width=0.95\linewidth]{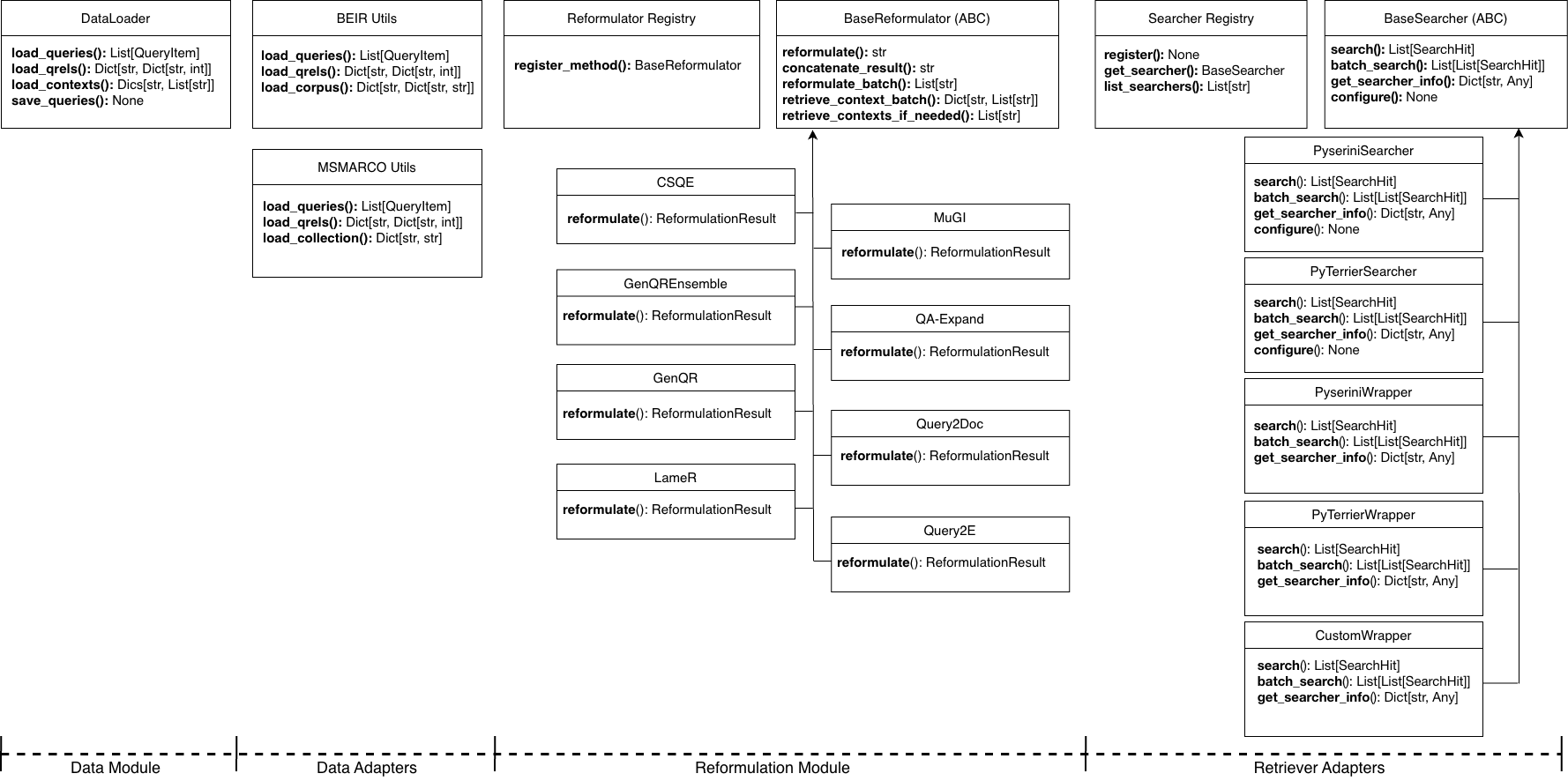}
\caption{Inheritance hierarchy for the main classes in the \texttt{QueryGym} Python package.}
\label{fig:class_diagram}
\vspace{-1em}
\end{figure*}

To address these challenges, we propose \texttt{QueryGym}\footnote{\url{https://querygym.readthedocs.io/}}, a well-structured, extensible, and publicly available toolkit designed to support research on LLM-based query reformulation. 
\texttt{QueryGym} is designed to facilitate the development of LLM-based query expansion strategies within a unified software framework. At its core, the toolkit is built around four key capabilities:  
(1) \textit{a unified reformulation framework} for standardizing the implementation of methods;  
(2) \textit{a retrieval-agnostic interface} for seamless integration with diverse IR retrieval libraries;  
(3) \textit{a centralized prompt bank} for reproducible prompt engineering and template management; and  
(4) \textit{LLM compatibility and reproducibility support} to enable implementation across diverse LLMs and prompting strategies.

At the center of \texttt{QueryGym} is the unified reformulation framework, which provides a standardized execution flow for implementing reformulation methods, managing prompts, interacting with LLMs, and formatting outputs. The toolkit supports batch reformulation with flexible concatenation strategies and robust handling of inputs and outputs. It also offers native support for popular IR datasets, including MS MARCO and BEIR, while also supporting custom and local formats through flexible loaders. 

\texttt{QueryGym} also leverages a retrieval-agnostic interface that enables seamless integration with diverse IR pipelines, such as Pyserini \cite{pyserini} and PyTerrier \cite{pyterrier}. This design ensures that query reformulation process can be performed through a standardized retrieval setting without requiring any pipeline reimplementation.

To support prompt experimentation, \texttt{QueryGym} introduces a centralized Prompt Bank, which manages versioned templates along with structured metadata. This enables prompt sharing and reuse across models and datasets and ensures reproducibility and transparency in prompt design. Importantly, the toolkit is fully LLM-compatible and supports both open-source models and API-based LLMs that can be accessed through OpenAI-compatible endpoints. These makes \texttt{QueryGym} a suitable toolkit to assess the impact of various LLMs and prompt variations.

Finally, the toolkit is built to help with reproducibility and scale. All experiments are driven by structured configuration files that control prompt version, model parameters, and retrieval settings. Reformulated queries are saved in both retrieval-ready and structured formats, while a CLI and high-level Python API provide flexible entry points for rapid prototyping and large-scale batch runs. Metadata, including prompt identifiers, LLM generation traces, and configuration parameters, is automatically logged to ensure that experimental results can be reliably audited and reproduced.

In our demo presentation, we will present \texttt{QueryGym} as a principled, flexible, and reproducible toolkit for LLM-based query reformulation. We (1) introduce the reformulation methods supported by \texttt{QueryGym} and illustrate how different methods can be employed; (2) demonstrate how new prompting methods and LLMs can be seamlessly integrated into the framework; (3) show how the unified interface enables fair, reproducible experimentation across datasets and retrieval backends; and (4) walk through real-world usage scenarios using the provided CLI and API, highlighting \texttt{QueryGym}’s value for both fast experimentation and scalable pipeline deployment.

\section{Toolkit Overview}

\texttt{QueryGym} is a modular and extensible Python toolkit designed to facilitate systematic research and development in LLM-based query reformulation. It can be easily installed via "\texttt{pip install querygym}". The toolkit is organized into four loosely coupled modules including \textit{Data Module and Data Adapters}, \textit{Reformulation Module}, \textit{Retriever Adapters}, and \textit{Configuration Utilities}. It exposes a consistent object-oriented interface for defining reformulation strategies, loading benchmark datasets, integrating with retrieval engines, and managing input/output workflows. Figure~\ref{fig:class_diagram} presents an overview of \texttt{QueryGym} core classes and their relationships.

\noindent\textbf{Data Module and Data Adapters.}  
The data module provides a lightweight, dependency-free interface for ingesting benchmark datasets. At its core is the \texttt{DataLoader} class, which handles the loading and saving of queries, qrels, and context passages in TSV or JSONL formats. To streamline experimentation on widely used benchmarks, \texttt{QueryGym} includes specialized adapters for datasets like BEIR~\cite{thakur2021beir} and MS MARCO~\cite{nguyen2016ms}, with utilities tailored to their respective file structures. This abstraction simplifies switching between datasets and ensures compatibility with retrieval pipelines and experimentation frameworks. The modular design also allows additional dataset adapters to be added with minimal effort, enabling broader applicability across IR benchmarks.

\noindent\textbf{Reformulation Module.}  
At the core of \texttt{QueryGym} is a unified reformulation framework centered on an abstract base class that standardizes method execution and encapsulates shared functionality for prompt rendering, context retrieval, and output formatting. The toolkit includes implementations of recent LLM-based reformulation methods drawn from literature, including Query2Doc~\cite{query2doc}, GenQR~\cite{Genqr}, GenQREnsemble~\cite{Genqrensemble}, MuGI~\cite{mugi}, QA-Expand~\cite{qa-expand}, LameR~\cite{LameR}, Query2E~\cite{exploringresearch}, and CSQE~\cite{csqe}. Each method is implemented as a modular subclass, adhering to a common interface, enabling consistent usage, extensibility, and integration.

To support extensibility, \texttt{QueryGym} adopts a lightweight decorator-based registration mechanism that allows new methods to be seamlessly added without altering the core framework. Once registered, these methods are immediately accessible through the toolkit’s Python API and CLI. This design facilitates rapid experimentation with novel prompting techniques while ensuring compatibility with the surrounding infrastructure for dataset management and retrieval integration.

\noindent\textbf{Retriever Integration and Wrappers.}  
For query reformulation methods that rely on retrieval context, such as those requiring passages or pseudo-relevance feedback, \texttt{QueryGym} defines a dedicated retriever abstraction through the \texttt{BaseSearcher} interface. This interface decouples reformulation logic from backend retrieval, supporting both single-query and batch retrieval workflows. It is compatible with widely-used IR toolkits such as Pyserini~\cite{pyserini} and PyTerrier~\cite{pyterrier}, which are integrated via dedicated wrappers.

Searchers can be instantiated dynamically through a registry-based mechanism, allowing flexible configuration at runtime. Provided wrappers, including support for Pyserini, PyTerrier, and custom search engines, enable seamless interoperability with existing IR tools. This retrieval-agnostic architecture ensures that reformulation methods can operate consistently across different retrieval backends without requiring changes to their core implementation.

\noindent\textbf{Prompt Management and Configuration.} Prompting logic is centralized through a YAML-based Prompt Bank, which stores template definitions along with metadata, role formatting (system/user/assistant), and version identifiers. Prompts are rendered dynamically with variable and are logged for full traceability. This enables systematic prompt tuning and sharing across experiments.

Overall, \texttt{QueryGym} has a clear separation between data handling, reformulation logic, retrieval integration, and prompt configuration, which allows researchers to experiment with new methods, prompts, and retrieval strategies without entangling components. By providing a consistent API, structured configuration, and interoperability with widely used IR toolkits, \texttt{QueryGym} lowers the barrier to reproducible experimentation and scalable deployment.

\vspace{-1em}
\section{Demonstrating Use Cases}

To illustrate the utility of \texttt{QueryGym}, we present several representative use cases that demonstrate how the toolkit facilitates query reformulation across different levels of complexity and integration. These examples highlight \texttt{QueryGym}'s suitability for both rapid prototyping and scalable experimentation in realistic setting.

\noindent\textbf{Basic Query Reformulation.}
Figure~\ref{fig:basic-usage} shows a basic example in which a set of user queries is reformulated using a single method and a specified LLM. This simple workflow is representative of lightweight experimentation, where researchers can iterate over prompting strategies and inspect reformulation outputs with minimal setup. The toolkit automatically handles batch processing, progress tracking, and result formatting. Results include both reformulated queries and method-specific metadata, enabling comprehensive downstream analysis and evaluation.

\begin{figure}[t]
\centering
\begin{minipage}{0.91\columnwidth} % ensures clean 
\begin{lstlisting}[style=ExactPython]
import querygym as qg
# Load queries
queries = qg.load_queries("examples/tiny_queries.tsv")
# Create reformulator
reformulator = qg.create_reformulator("query2doc", model="gpt-4")
# Reformulate
results = reformulator.reformulate_batch(queries)
# Show results
for r in results:
    print(f"{r.qid}: {r.original} -> {r.reformulated}")

\end{lstlisting}
\end{minipage}
\vspace{-1em}
\caption{Example usage of \texttt{QueryGym} for query reformulation.}
\vspace{-1em}
\label{fig:basic-usage}
\end{figure}

\noindent\textbf{Context-Based Reformulation with Retrieval.}
Query reformulation methods that rely on external context such as top-ranked passages can be directly integrated with retrieval engines using \texttt{QueryGym}’s retriever abstraction. Figure~\ref{fig:retrieval} illustrates an end-to-end pipeline where a context-aware method is applied to a benchmark dataset using a prebuilt Pyserini index. The retrieval system is seamlessly wrapped using \texttt{QueryGym}’s utility functions, allowing reformulation strategies to incorporate retrieved content without altering core logic or re-implementing IR components.

This retrieval-agnostic integration allows researchers to run retrieval-augmented prompting methods in realistic scenarios using widely adopted toolkits. \texttt{QueryGym}’s interface ensures compatibility with both batch and single-query workflows and supports configuration of retrieval parameters such as ranking models and passage cutoffs. By bridging LLM-based reformulation with established IR infrastructure, \texttt{QueryGym} enables reproducible experimentation within modular and extensible pipelines.

\begin{figure}[H]
\centering

\begin{minipage}{0.91\columnwidth} % ensures clean alignment in ACM columns
\begin{lstlisting}[style=ExactPython]
import querygym as qg
from pyserini.search.lucene import LuceneSearcher
# User configuration
DATA_DIR = "data" 
# Load queries
queries = qg.loaders.msmarco.load_queries(f"{DATA_DIR}/queries.tsv")
# Initialize searcher
searcher = LuceneSearcher.from_prebuilt_index('msmarco-v1-passage')
searcher.set_bm25(k1=0.9, b=0.4)
# Wrap searcher and create reformulator
wrapped_searcher = qg.wrap_pyserini_searcher(searcher, answer_key="contents")
reformulator = qg.create_reformulator(
    method_name="csqe",
    model="gpt-4",
    params={
        "searcher": wrapped_searcher,
        "retrieval_k": 10,
        "gen_passages": 5,
    },
    llm_config={
        "temperature": 1.0,
        "max_tokens": 128,
    }
)
# Reformulate
results = reformulator.reformulate_batch(queries)
\end{lstlisting}
\end{minipage}

\caption{Integrated pipeline for Pyserini retrieval and \texttt{QueryGym} reformulation.}

\label{fig:retrieval}
\end{figure}

\noindent\textbf{Leaderboard and Benchmarking.}
\texttt{QueryGym} facilitates reproducible method comparison across datasets under controlled experimental conditions. Figure~\ref{fig:benchmarking} demonstrates a systematic experimentation pipeline that benchmarks six reformulation methods on three MS MARCO datasets with identical LLM configurations.

The pipeline enforces identical experimental configurations across all runs through centralized parameter management. Method-specific requirements are handled programmatically, while outputs are organized hierarchically for immediate downstream evaluation with standard IR tools.

This workflow illustrates how \texttt{QueryGym} facilitates systematic IR research, enabling researchers to scale from single-method to multi-method, multi-dataset experimentation while preserving reproducibility and methodological consistency.

\begin{figure}[t]
\centering

\begin{minipage}{0.91\columnwidth} % ensures clean 
\begin{lstlisting}[style=ExactPython]
import querygym as qg
from pyserini.search.lucene import LuceneSearcher
# Benchmark configuration
DATASETS = ["dev_small", "trecdl2019", "trecdl2020"]
METHODS = ["genqr", "genqr_ensemble", "query2doc", "qa_expand", "lamer", "csqe"]
CONTEXT_METHODS = {"lamer", "csqe"}
DATA_DIR = "data"
LLM_CONFIG = {
    "model": "gpt-4o",
    "temperature": 0.8,
    "max_tokens": 256,
    "seed": 42
}
# Initialize shared index for all MS MARCO datasets
searcher = LuceneSearcher.from_prebuilt_index('msmarco-v1-passage')
searcher.set_bm25(k1=0.9, b=0.4)
wrapped_searcher = qg.wrap_pyserini_searcher(searcher, answer_key="contents")
for dataset in DATASETS:
    queries = qg.loaders.msmarco.load_queries(f"{DATA_DIR}/{dataset}/queries.tsv")
    for method in METHODS:
        method_params = (
            {"searcher": wrapped_searcher, "retrieval_k": 10}
            if method in CONTEXT_METHODS else {}
        )
        reformulator = qg.create_reformulator(
            method,
            model=LLM_CONFIG["model"],
            params=method_params,
            llm_config={k: v for k, v in LLM_CONFIG.items()         if k != "model"},
            seed=LLM_CONFIG["seed"]
        )
        reformulated = reformulator.reformulate_batch(queries)
        # Export reformulated queries
        qg.DataLoader.save_queries(
            [qg.QueryItem(r.qid, r.reformulated) for r in           reformulated],
            f"benchmark/{dataset}/{method}.tsv"
        )
\end{lstlisting}
\end{minipage}

\caption{Multi-method benchmarking pipeline across datasets under controlled conditions.}

\label{fig:benchmarking}
\end{figure}

\section{Concluding Remarks}

\texttt{QueryGym} provides a cohesive and practical environment for investigating LLM-based query reformulation, designed to facilitate extensibility, controlled experimentation, and seamless interoperability with modern retrieval pipelines. The demo will guide attendees through the core capabilities of the toolkit and illustrate how its unified abstractions simplify the development of reformulation strategies. Our demonstration will showcase the following:

\begin{itemize}[leftmargin=1em]
\item interactive execution of reformulation methods via both the Python API and command-line interface, highlighting the ease of iterating over models, prompts, and reformulation strategies;
\item end-to-end workflows that integrate reformulation with retrieval tools such as Pyserini and PyTerrier, illustrating retrieval-agnostic design choices and straightforward backend substitution;
\item prompt selection, versioning, and metadata inspection enabled by the centralized prompt bank, demonstrating how prompt management supports transparent and reproducible experimentation;
\item benchmarking pipelines that compare multiple reformulation methods across datasets under consistent settings, enabling fair and repeatable experimentation;
\item incorporation of a new reformulation method or LLM backend to demonstrate the lightweight extensibility mechanism and the clarity of the underlying abstractions;

\end{itemize}

Through these examples, the demonstration aims to show how \texttt{QueryGym} enables both rapid prototyping and systematic experimentation of LLM-driven query reformulation, providing a reproducible, modular, and extensible foundation for future research. The complete open-source implementation of \texttt{QueryGym} is available at \url{https://github.com/radinhamidi/QueryGym}.

\bibliographystyle{ACM-Reference-Format}
\balance
\bibliography{references} 

\end{document}